\documentclass[aps,prb,reprint,superscriptaddress,showpacs,citeautoscript]{revtex4-1}

\usepackage{graphicx}
\usepackage{color}

\usepackage{amsmath}
\usepackage{accents}

\makeatletter
\def\widebar{\accentset{{\cc@style\underline{\mskip10mu}}}}
\makeatother

\begin{document}

\title{Glass-like features of crystalline solids in the quantum critical regime}


\author{Y. Ishii}
\email{ishii@mtr.osakafu-u.ac.jp}
\affiliation{Department of Materials Science, Osaka Prefecture University, Sakai, Osaka 599-8531, Japan.}

\author{Y. Ouchi}
\affiliation{Department of Materials Science, Osaka Prefecture University, Sakai, Osaka 599-8531, Japan.}

\author{S. Kawaguchi}
\affiliation{Japan Synchrotron Radiation Research Institute (JASRI), SPring-8, Sayo, Hyogo 679-5198, Japan.}

\author{H. Ishibashi}
%\email{kawaguchi@spring8.or.jp}
%\thanks{}
\affiliation{Department of Physical Science, Osaka Prefecture University, Sakai, Osaka 599-8531, Japan.}

\author{Y. Kubota}
\affiliation{Department of Physical Science, Osaka Prefecture University, Sakai, Osaka 599-8531, Japan.}

\author{S. Mori}
\affiliation{Department of Materials Science, Osaka Prefecture University, Sakai, Osaka 599-8531, Japan.}

%\author{Y. Kuroiwa}
%\affiliation{Department of Physics, Hiroshima University, Higashi-Hiroshima, Hiroshima 739-8526, Japan.}


\date{\today}

\begin{abstract}

There has been growing interest in structural quantum phase transitions and quantum fluctuations of phonons in the research area of condensed matter physics.
Here, we report the observation of glass-like features in the lattice heat capacity of a stuffed tridymite-type crystal, Ba$_{1-x}$Sr$_x$Al$_2$O$_4$, a candidate compound of quantum paraelectrics.
Substitutional chemical suppression of the ferroelectric phase transition temperature ($T_{\rm C}$) of Ba$_{1-x}$Sr$_x$Al$_2$O$_4$ results in the disappearance of the $T_{\rm C}$ at $x=0.07$.
For the compositional window of $x=0.2$--0.5, the lattice heat capacity is enhanced below approximately 10 K and diverges from the $T^3$-scaling law below 2.5 K.
Synchrotron X-ray diffraction experiments on single crystals reveal the weakly correlated disorder in the crystal structure that survives down to low temperature; this disorder is responsible for the observed excess heat capacity.
These features can be considered one of the manifestations of structural quantum fluctuations.

\end{abstract}

\maketitle
%\section*{Main text}

\section{Introduction}

Quantum criticality has long been a major subject in condensed matter physics.
For decades, a considerable amount of effort has been devoted to strongly correlated electron systems and has provided us with accumulative knowledge of quantum fluctuations in spin systems.
Very recently, interests in the quantum criticality of phonons as a counterpart of spins, such as quantum paraelectrics\cite{Rowley_NatPhys} or ferroelectric superconductivity\cite{Edge_PRL115,Ferroelectric_Super}, has grown rapidly.
However, there have been limited works that address an essential issue: what is the quantum critical behaviour of a phonon?

Ferroelectric phase transitions are usually caused by condensation of a soft mode and a subsequent structural phase transition.
The resultant low-temperature phase can be regarded as an ordered phase of an associated soft mode.
In so-called quantum paraelectrics, including several perovskite-type oxides\cite{STO_quantumpara,KTO_quantumpara,Taniguchi_CdCaTiO3} and organic compounds\cite{Horiuchi229,Rowley_organic,Horiuchi_NatComm}, the ferroelectric phase transition is suppressed down to absolute zero temperature by tuning parameters such as chemical composition or external pressure, and dielectric constants remarkably increase upon cooling without a ferroelectric phase transition.
According to Raman scattering experiments\cite{TaniguchiPRL} on $^{18}$O-exchanged SrTiO$_3$, the soft mode does not condense down to extremely low temperatures, indicating that the low-energy phonons survive even at absolute zero temperature.
However, despite a large number of studies, including other spectroscopic experiments such as NMR\cite{NMR_STO18} and second-harmonic generation (SHG)\cite{SHG_STO18}, the nature of ``structural'' quantum fluctuation is unclear owing to the extremely high dielectric constants, e.g., $\approx$20,000, at low temperature, which cause the system to become highly sensitive to external perturbation.

\begin{figure*}[t]
\begin{center}
\includegraphics[width=155mm]{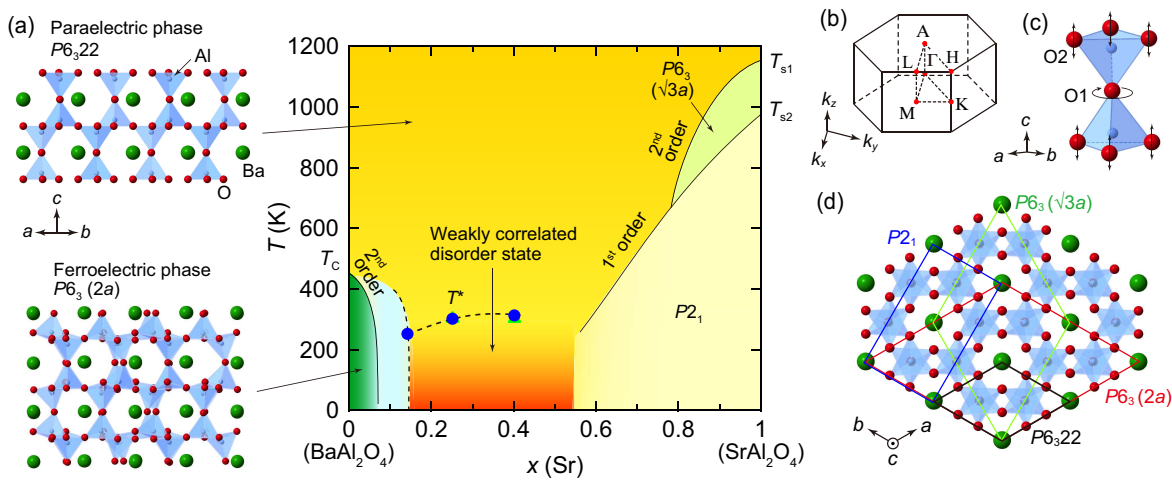}% Here is how to import EPS art
\caption{\label{PhaseDiagram} {\bf Structural phase transitions and weakly correlated disorder state of  Ba$_{1-x}$Sr$_x$Al$_2$O$_4$.} (a) Phase diagram of Ba$_{1-x}$Sr$_x$Al$_2$O$_4$\cite{Ishii_PRB93,Ishii_PRB94,Kawaguchi_PRB,Ishii_SSC249}. 
In the Ba-rich region, condensation of a soft mode characterized as M$_2$ occurs at the ferroelectric transition temperature ($T_{\rm C}$) and forms a $P6_3$ phase (2$a$ structure).
The $T_{\rm C}$ is rapidly suppressed by increasing $x$.
At $x=0.07$, the border of ferroelectricity, the three distinct instabilities, namely, those at the M and K points and along the $\Gamma$-A line in reciprocal space, emerge simultaneously and form superstructures.
The $\sqrt{3}a$ structure, which is the ordered phase of the K$_2$ mode, exists in a narrow compositional window shaded light blue.
The $T^*$ values for the M and K points are indicated by closed blue circles and green triangles, respectively.
In the region below $T^*$, weakly correlated structural disorder is dominant. 
 (b) The first Brillouin zone of the hexagonal crystal. (c) The atomic displacement patterns commonly observed in the M$_2$ and K$_2$ modes\cite{Ishii_PRB93}. (d) Various cell settings of Ba$_{1-x}$Sr$_x$Al$_2$O$_4$.}
\end{center}
\end{figure*}

This study highlights an improper ferroelectric, BaAl$_2$O$_4$\cite{Stokes_improper}, and a Sr-substituted one that comprise a network structure of AlO$_4$ tetrahedra with shared vertices.
The phase diagram of Ba$_{1-x}$Sr$_x$Al$_2$O$_4$ is shown in Fig. \ref{PhaseDiagram} (a).
These materials crystallize in the same structure of space group $P6_322$ at high temperature over the whole compositional range.
The high-temperature phase of BaAl$_2$O$_4$ possesses two acoustic soft modes with nearly the same instability\cite{Ishii_PRB93}, considered M$_2$ and K$_2$ in irreducible representation, which are characterized as collective tilting with significant vibration of oxygen atoms in the AlO$_4$ network.
Fig. \ref{PhaseDiagram} (c) represents the atomic displacement patterns obtained by using first-principle calculations, which are commonly observed for both modes, M$_2$ and K$_2$.
The O1 atoms, which link the AlO$_4$ along the $c$ axis, vibrate with circular motion in the $ab$ plane around the 3-fold axis.
Because of this significant vibration, the O1 atoms exhibit anomalously large isotropic thermal factors in synchrotron X-ray structural refinements\cite{Kawaguchi_PRB}.
Interestingly, the $T_{\rm C}$ is strongly suppressed by a small amount of Sr substitution for Ba and disappears at $x=0.07$\cite{Ishii_PRB94}, as shown in Fig. \ref{PhaseDiagram} (a).
Outside the ferroelectric-paraelectric phase boundary, the isotropic thermal factor of the O1 atom exhibits an unusually large value, which is largely independent of temperature down to low temperature.
This fact motivated us to investigate this compositional window in detail in expectation of observing a structurally fluctuating state lying there.

This paper reports that the Ba$_{1-x}$Sr$_x$Al$_2$O$_4$ crystal exhibits glass-like features in terms of its lattice heat capacity, which are ascribed to the weakly correlated disorder in the atomic arrangement.
These features can be considered a possible manifestation of the structural quantum fluctuation.

\begin{figure}[t]
\begin{center}
\includegraphics[width=75mm]{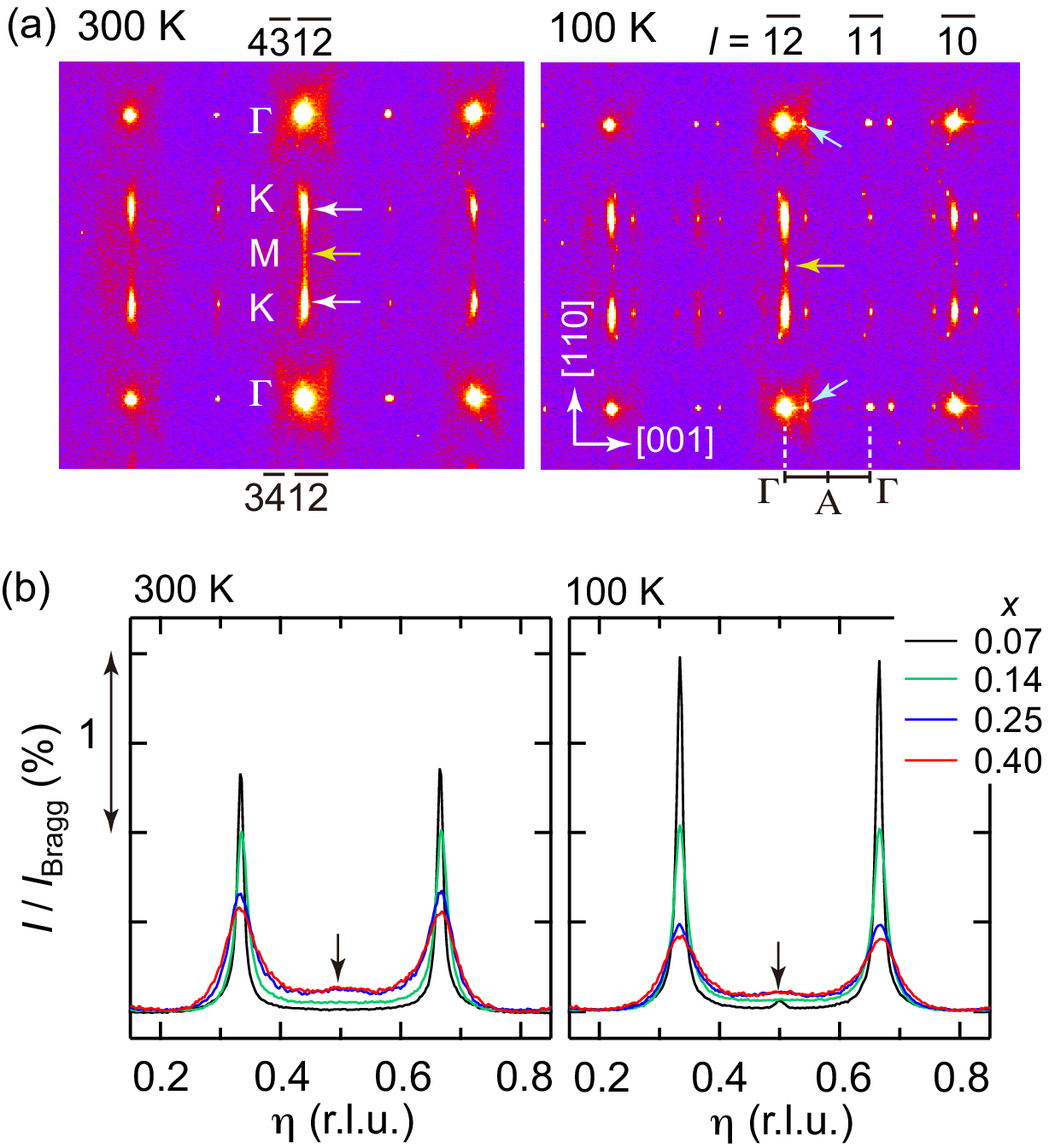}% Here is how to import EPS art
\caption{\label{Xray} {\bf Superlattice reflections and diffuse scatterings of Ba$_{1-x}$Sr$_x$Al$_2$O$_4$.}
(a) Synchrotron X-ray diffraction patterns at 300 K (left) and 100 K (right) of a Ba$_{1-x}$Sr$_x$Al$_2$O$_4$ ($x=0.07$) single crystal. 
The white arrows indicate the superlattice reflections at the K point.
At 100 K, satellite reflections are visible along the $\Gamma-A$ line, as marked by blue arrows.
The diffuse scattering at the M point, which is observed at 300 K, becomes a superlattice reflection at 100 K, as marked by yellow arrows.
The weak reflections found just above and below the K points come from surface microcrystallites and therefore are not important.
(b) Integrated intensity profiles of $x=0.07$--0.40 plotted along the [110] direction at 300 K (left) and 100 K (right). $\eta=0$ and 1 correspond to reciprocal points of the lower and upper fundamental reflections, respectively.
The profiles are shifted along the vertical axis for comparison.
At both temperatures, the superstructure at the K points observed for $x=0.07$ is gradually suppressed as $x$ increases.
In contrast, the diffuse-scattering intensity at the M point increases, as indicated by arrows.}
\end{center}
\end{figure}

\section{Experimental}

Single crystals of Ba$_{1-x}$Sr$_x$Al$_2$O$_4$ were grown by the self-flux method. Previously prepared Ba$_{1-x}$Sr$_x$Al$_2$O$_4$ ($x=0.2$, 0.4, 0.6, 0.8) and BaCO$_3$ powders were mixed at a molar ratio of 6.7--8.8:3. Ba$_{1-x}$Sr$_x$Al$_2$O$_4$ powder was prepared by using a conventional solid-state reaction. The mixture was placed in a platinum crucible. After heating at 1650$^{\circ}$C for 6 h, the crucible was slowly cooled to 1440--1350$^{\circ}$C at a rate of 2$^{\circ}$C/h and then cooled in a furnace to room temperature. 
The shiny, colourless crystals had hexagonal-shaped edges of approximately 50--100 $\mu$m in length and were mechanically separated from the flux.

The Sr concentrations of the obtained crystals were found to be $x=0.07$, 0.14, 0.25, and 0.40 by means of the inductively coupled plasma method.
Synchrotron X-ray thermal diffuse scattering was performed for these single crystals over a temperature range of 100--400 K at the BL02B1 beamline of SPring-8. The incident X-ray radiation was set at 25 keV. The diffraction intensities were recorded on a large cylindrical image-plate camera\cite{IPcamera}. 
Temperature control was performed using N$_2$ gas flow.

Heat capacity was measured for the polycrystalline samples of $x=0$--0.5 using a heat-relaxation method in a physical property measurement system (PPMS, Quantum Design).
For the measurements, the powder sample was uniaxially pressed into a pellet and then pressed again under hydrostatic pressure.
The samples were heated at 1450$^{\circ}$C for 48 h and cut into rectangular shapes.
Typical sample thickness, mass, and density were 0.35 mm, 7 mg, and 85--90\% of the theoretical values, respectively.

%%%%%%%%%%%%%%%%%%%%%%%%%%%%%%%%%%%%%%%%%%
%\section{Results and Discussion}
%%%%%%%%%%%%%%%%%%%%%%%%%%%%%%%%%%%%%%%%%%

\section{Results and Discussion}

%\vspace{3mm}
%\noindent {\bf Structural phase transitions}

In BaAl$_2$O$_4$, the K$_2$ and M$_2$ modes soften simultaneously at $T_{\rm C}$ = 450 K because of competing instability\cite{Ishii_PRB93}.
However, the K$_2$ mode is electrostatically unfavourable due to the larger distortion in the AlO$_4$ tetrahedra in this mode than in the M$_2$ mode.
As a result, the M$_2$ mode eventually condenses to form the corresponding low-temperature phase with a cell volume of $2a \times 2b \times c$ ($P6_3$, $2a$ structure), in which the collective vibration of AlO$_4$ tetrahedra is frozen.
Both low-temperature phases, $P6_3$ (2$a$) and $P2_1$, are caused by the condensation of the M$_2$ mode\cite{Stokes_improper,Perez-Mato-PRB79}; the $2a$ type is the structure where the M$_2$ mode condenses in all of the three equivalent $\langle110\rangle$ directions ($3q$), whereas the $P2_1$ phase is the structure where it condenses in only one ($1q$) of the three directions.
The condensation of the K$_2$ mode is known to form the $\sqrt{3}a$ structure\cite{Rodehorst}.
The various cell settings are depicted in Fig. \ref{PhaseDiagram} (d).
%The $P2_1$ phase found in the Sr-rich side is another condensed state of the M$_2$ mode.

We unexpectedly observed a peculiar formation of superstructures on the verge of the ferroelectric phase.
Fig. \ref{Xray}(a) shows the synchrotron X-ray diffraction pattern of the $x=0.07$ single crystal near the 4$\widebar{3}\widebar{12}$ and 3$\widebar{4}\widebar{12}$ fundamental reflections at 300 K and 100 K. 
Strong superlattice reflections are observed at the K point at both temperatures, as marked by white arrows.
According to U. Rodehorst {\it et al.}\cite{Rodehorst}, these superlattice reflections come from the $\sqrt{3}a$ structure.
In addition, weak diffuse scattering is observed at the M point at 300 K and becomes the weak superlattice reflection of the $2a$ structure at 100 K, as marked by yellow arrows.
Furthermore, satellite reflections develop along the [001] at 100 K, as marked by blue arrows in Fig. \ref{Xray}(a, right).
This satellite reflection is probably caused by the tertiary instability, which has been observed along the $\Gamma-A$ line in our previous phonon calculations \cite{Ishii_PRB93}.
The M-point superlattice reflection and the satellite reflections were not observed at $x\geq0.14$.
In other words, all of the three instabilities at K, M, and $\Gamma-A$ emerge on the border of ferroelectricity.

\begin{figure}[t]
\begin{center}
\includegraphics[width=85mm]{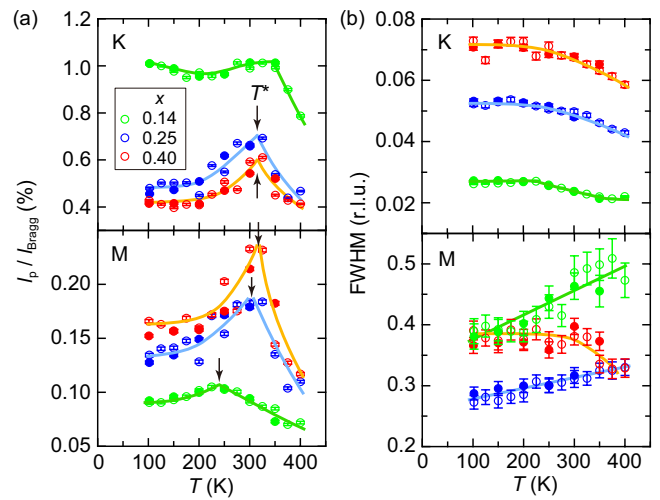}% Here is how to import EPS art
\caption{\label{XRDSummary} {\bf Temperature variation in the peak intensity and FWHM.}
(a) Normalized peak intensity ($I_{\rm p}/I_{\rm Bragg}$) at the K (top) and M points (bottom) plotted against temperature. 
Except for the $I_{\rm p}/I_{\rm Bragg}$ at the K point of the $x=0.14$ crystal, each plot shows a maximum at approximately 250--300 K, as denoted by $T^*$, indicating incomplete softening of the K$_2$ and M$_2$ modes.
Solid lines are guides for eyes.
The open and closed symbols represent the data obtained upon heating and cooling, respectively.
(b) The FWHM of the peaks observed at the K (top) and M (bottom) points.}
\end{center}
\end{figure}

%\vspace{3mm}
%\noindent {\bf Weakly correlated disorder state}

The superlattice reflections at the K point were observed in a narrow compositional window of $x=0.07$ to 0.14 below 400 K, which was the upper temperature limit of the present study.
Notably, these reflections are significantly elongated along the [110], as shown in Fig. \ref{Xray}(a).
Fig. \ref{Xray}(b) shows the intensity profiles plotted along [110] between two fundamental reflections obtained at 300 K and 100 K for crystals of $x=0.07$, 0.14, 0.25 and 0.40.
$\eta=0$ and $\eta=1$ correspond to the reciprocal points of the lower and upper fundamental reflections, respectively.
The profiles are normalized so that the fundamental reflection at $\eta=0$ is equal in intensity.
%by using the lower fundamental reflection of each composition.
%Backgrounds were subtracted from the data.
The profiles are shifted along the vertical axis for comparison.
As $x$ increases, the superlattice intensity at the K point gradually decreases, the peak width becomes broader, and the reflection transforms into diffuse scattering at $x=0.25$.
That is, the long-range correlation of the $\sqrt{3}a$ structure is gradually reduced, and the system enters a structurally disordered state. %, which is indicated by green in Fig. \ref{PhaseDiagram}.
In this state, one can also find diffuse scattering at the M point, as indicated by arrows in Fig. \ref{Xray}(b), whose intensity increases as $x$ increases after suppression of the condensation of the M$_2$ mode.

The intensity profiles shown in Fig. \ref{Xray}(b) can be decomposed by using Lorentzian functions.
Details of the fitting procedures are shown in Supplementary information.
The maximum intensity of the peak ($I_{\rm p} / I_{\rm Bragg}$) and the full width at half maximum (FWHM) are plotted against temperature in Figs. \ref{XRDSummary}(a) and (b), respectively.
% which is indicated in Fig. \ref{Xray}(b, left)
Fig. \ref{XRDSummary}(a, top) shows the temperature dependence of $I_{\rm p} / I_{\rm Bragg}$ at the K point.
For $x=0.25$ and 0.40, each intensity reaches a maximum near 300 K, as denoted by $T^*$, and shows weaker temperature dependence below $\approx$200 K.
This trend is also observed at the M point, as shown in Fig. \ref{XRDSummary}(a, bottom), whose diffuse scattering intensity shows a maximum at 250, 300, and 310 K for $x=0.14$, 0.25 and 0.40, respectively.
Although the $I_{\rm p} / I_{\rm Bragg}$ at the K point decreases as $x$ increases over the whole temperature range, which corresponds to suppression of the K$_2$ mode, the $I_{\rm p} / I_{\rm Bragg}$ at the M point is enhanced as $x$ increases.

The top and bottom panels of Fig. \ref{XRDSummary}(b) show the FWHM for the K and M points, respectively.
As shown in Fig. \ref{XRDSummary}(b, top), the FWHM for the K point systematically increases as $x$ increases, which is reasonably ascribed to the suppression of the K$_2$ mode.
As the temperature decreases, the FWHM values show an increasing trend and saturate at $\approx$200 K, below which the $I_{\rm p} / I_{\rm Bragg}$ becomes nearly constant.
In contrast to this increasing trend at the K point, the FWHM values at the M point for $x=0.14$ and 0.25 decrease as the temperature decreases, as shown in Fig. \ref{XRDSummary}(b, bottom), which indicates that the correlation of the M$_2$ mode is enhanced upon cooling.
On the other hand, the FWHM of $x=0.40$ exhibits a similar trend as the K point.

Although Sr substitution would slightly affect the phonon dispersion,
it would not fundamentally change the essential character of the soft modes.
The enhancement in the peak intensity at $T^*$ observed in Fig. \ref{XRDSummary} (a) is simply attributable to the reduction in the phonon frequencies, $\omega$, because the thermal diffuse scattering intensity caused by phonon scales as $\omega^{-2}$.
That is, both modes reduce the phonon frequency towards $T^*$, although they cannot condense completely.
This effect results in a characteristic state of weakly correlated disorder.
Two possible pictures can be proposed: 
the first is the soft modes that survive in this state as dynamic disorder, and the second is the glass-like short-range order as static disorder caused by the incomplete condensation of the soft modes.

\begin{figure}[t]
\begin{center}
\includegraphics[width=85mm]{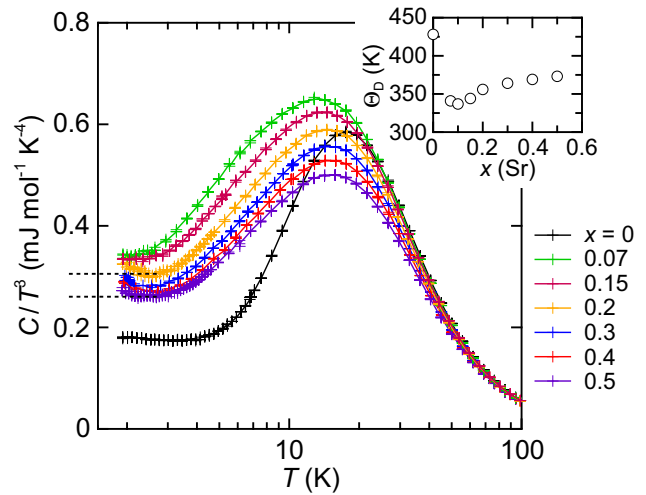}% Here is how to import EPS art
\caption{\label{HeatCapacity} {\bf Glass-like features in the lattice heat capacity of a Ba$_{1-x}$Sr$_x$Al$_2$O$_4$ crystal.} 
The lattice heat capacity ($C_p$) divided by $T^3$ is plotted against temperature.
The $x=0.2$--0.5 samples show an enhanced lattice heat capacity near 10 K and an upturn below $\approx$2.5 K, which are typical behaviours in amorphous solids. 
Inset shows the Debye temperature ($\Theta_{\rm D}$) obtained by using the coefficient of $C/T^3$. 
For $x=0.2$--0.5, the coefficient of $C/T^3$ was determined by using the local minimum value, as indicated by broken lines.}
\end{center}
\end{figure}

%\vspace{3mm}
%\noindent {\bf Glass-like features in the quantum critical regime}

Fig. \ref{HeatCapacity} shows the heat capacity divided by $T^3$ plotted against $T$ on a logarithmic scale.
Polycrystalline samples of $x=0$--0.5 were used for the measurements.
Special care was taken with regard to sample thickness and density.
The $C/T^3$ of $x=0$ follows the $T^3$-scaling law and shows a constant value below 4 K.
The Debye temperature ($\Theta_{\rm D}$) is evaluated to be 428 K from the coefficient.
This $T^3$-scaling law is true up to $x=0.15$, below which the long-range order of the 2$a$ or $\sqrt{3}a$ structure is found at low temperature.
The $x=1$ crystal, the end material of the Sr-rich side, also follows this classic law.
However, for $x=0.2$--0.5 crystals, the heat capacity surprisingly diverges from this scaling law below 2.5--3 K, as shown in Fig. \ref{HeatCapacity}.
In addition, a large enhancement in $C/T^3$ is observed below 10 K.

These two features are very similar to those observed in amorphous solids, such as amorphous SiO$_2$.
That is, the lattice heat capacity of amorphous solids varies as $c_1T+c_2T^3$ below 1 K and shows a hump of $C/T^3$ at approximately 10 K\cite{GlassCp_PRB1971,Glass-Cp_PRL,MERTIG1984369,QueenPRL}.
Although the mechanism is still under debate\cite{BosonPeak_SSC, BosonPeak_Nakayama,BP_Localized, BosonPeak_vanHoveSingularity, LocalFavoredStructure, AcousticMode_PRL1, AcousticMode_PRL2,Buchenau_PRB43,Parshin_PRB}, the excess heat capacity observed in the amorphous system has generally been accepted to arise from the enhanced density of state\cite{Buchenau_PRL, Buchenau_PRB} of acoustic phonons caused by the disordered structure\cite{AcousticMode_NatMater,Chumakov_PRL}.
Note that the M$_2$ and K$_2$ modes in BaAl$_2$O$_4$ are also acoustic modes.
Obviously, the glassy behaviour observed in the present system is attributable to the disordered state found below $T^*$ outside the ordered state, although it is unclear whether the disordered state is dynamic or static.

Excess heat capacity has also been observed in relaxor ferroelectrics, which comprise short-range ordered domains called polar nano-regions \cite{relaxor1, relaxor2, relaxor3}.
Relaxor ferroelectrics show a characteristic Burns temperature, below which polar and non-polar phase separation takes place at the nanoscale.
A soft mode is condensed in the polar nano-regions, and the dielectric constant shows a broad and large peak.
In the case of relaxor ferroelectrics, this structural inhomogeneity is considered the reason for the excess heat capacity\cite{relaxor2}.
Because the $T_{\rm C}$ disappears at $x=0.07$ in Ba$_{1-x}$Sr$_x$Al$_2$O$_4$, the phenomena observed in the present system are different from those observed in relaxor ferroelectrics.

The inset of Fig. \ref{HeatCapacity} shows the Debye temperature ($\Theta_{\rm D}$) plotted against $x$.
These values were obtained from the coefficients of $C/T^3$ for $x=0$--0.15.
For $x\geq0.2$, the local minimum values of $C/T^3$ were used for the calculation, as shown by broken lines in Fig. \ref{HeatCapacity}.
$\Theta_{\rm D}$ exhibits a valley near $x=0.1$, indicating that the Debye frequency rapidly decreases towards the ferroelectric-paraelectric phase boundary.
%is close to the phase-boundary composition,
If we devise a concept, an ``effective" mass of vibrating atoms that should affect the phonon frequency, this behaviour may correspond to the enhanced effective mass of electrons in spin systems near the quantum critical point (QCP).
%If this rapid reduction of the phonon frequency has a relevance to "effective" mass of vibrating atoms, 
%This means that the lattice vibration softens near the ferroelectric-paraelectric phase boundary.
Above $x$=0.1, it in turn increases towards the composition where the $P2_1$ phase emerges.
% above $x$=0.6.

%\vspace{3mm}
%\noindent {\bf Quantum critical phenomena in phonon systems}

The above results can be summarized as follows.
The $\sqrt{3}a$ structure exists in a narrow compositional window right next to the ferroelectric $2a$ phase.
This $\sqrt{3}a$ structure is the condensed state of the K$_2$ mode.
Because the condensation of the M$_2$ mode is suppressed outside the ferroelectric-paraelectric phase boundary, the secondary instability at the K point temporarily appears.
Simultaneously, the tertiary instability along $\Gamma-A$ also emerges.
However, the condensed state of the K$_2$ mode is inherently unfavourable in terms of Coulomb energy.
In addition, the instability at $\Gamma-A$ is very small\cite{Ishii_PRB93}.
For these reasons, the temporary emerged states are easily destroyed by a further increase in $x$.
In the fluctuation state lying outside the ordered state of the K$_2$ mode, the lattice heat capacity exhibits glassy behaviour.
While the K$_2$ mode is suppressed in this fluctuation state, the M$_2$ mode is slightly enhanced as $x$ increases.
Considering that the substitution of Ba with Sr ions having a smaller ionic size results in a wider space for the AlO$_4$ vibration, it is reasonable to think that the lattice flexibility is enhanced as $x$ increases.
That is, the M$_2$ mode is further enhanced in this fluctuation state, but it cannot develop a strong correlation even at low temperature due to this enhanced lattice flexibility.
A further increase in $x$ turns into the $P2_1$ phase in the Sr-rich side, which is another condensed form of the M$_2$ mode\cite{Perez-Mato-PRB79}.

In fact, the ordered state of the K$_2$ mode has not been clearly observed in powder samples. 
In our previous electron diffraction experiments using powder samples, the diffuse scatterings at the K points disappeared for all compositions of $x=0.07$--0.5 at approximately 200 K\cite{Ishii_SciRep,Ishii_PRB94}, below which the M-point diffuse scatterings survive.
We have also performed electron diffraction experiments on the $x=0.14$ single crystal by means of the crush method.
However, the scattering at the K point was diffusive and disappeared at 155 K upon cooling without condensation, below which only the diffuse scatterings at the M point were observed.
These findings imply that the condensation of the K$_2$ mode is strongly affected by extrinsic factors such as strain energy due to small grain size. 
The satellite reflections along $\Gamma-A$ for $x=0.07$ have also not been observed in our previous electron diffraction experiments using powder samples.

%%%%%%%%%%%%%%%%%%%%%%%cut
%%%%%%%%%%%%%%%%%%%%%%%
%In fact, the superlattice reflections at the K point are not observed in our previous studies using powder samples. The electron diffraction patterns for $x=0.06$ and 0.07 powder samples show relatively strong intensity at the K point. However, this intensity does not develop as sharp superlattice reflections as temperature decreases and disappears approximately at $\sim$200 K. For all compositions up to $x = 0.5$ powder samples, the scattering intensity at the K point is weak and diffusive and disappears at around also 200 K. The scattering at the M points remains diffusive and does not condense. In addition, our TEM experiments by crushing method applied for the $x=0.14$ single crystals have also revealed that the scattering at the K point is diffusive and disappears at 155 K on cooling without condensation, below which only the diffuse scatterings at the M point are observed. These things imply that the condensation of the K$_2$ mode strongly affected by the extrinsic factors, such as grain size and strain energy within the crystallite. The satellite reflections along $\Gamma-A$ for $x=0.07$ are also not observed in the electron diffraction experiments using powder samples.
%%%%%%%%%%%%%%%%%%%%%%%cut
%%%%%%%%%%%%%%%%%%%%%%%

Unlike the so-called quantum paraelectrics, the present system does not exhibit a divergent enhancement in the dielectric constant.
This result is probably because the polarization is the secondary order parameter in BaAl$_2$O$_4$, whereas it is the primary parameter in the so-called quantum paraelectrics. 
In addition, it is unclear where the QCP is in this material because of the existence of the ordered state of the K$_2$ mode.
%In addition, it is unclear where the QCP is in this material because the phase transition already vanishes at a fairly high temperature.
Nevertheless, the phase-boundary composition, $x=0.07$, seems to be important.
In this material, the rapid suppression of the ferroelectric phase gives rise to the fluctuation state in which the weakly correlated disorder is dominant and contributes to the additional lattice heat capacity near absolute zero temperature.
In $^{18}$O-exchanged SrTiO$_3$, a typical quantum paraelectric material, a soft mode has been revealed to survive down to absolute zero temperature outside the quantum critical point\cite{TaniguchiPRL}.
These two materials have a common feature in that the structural fluctuation is maintained down to absolute zero temperature in a disordered state.

Analogous to the ordered state of conventional spin systems, the structural phase transition associated with a soft mode is regarded as an ordering of phonons.
Generally, enhanced quantum fluctuations near the QCP in a spin system are responsible for anomalous states such as superconductivity.
The region outside the QCP, where electrons behave as a Fermi liquid, is disordered.
When we consider the corresponding disordered state of phonons in the sense of quantum criticality, it would be the structurally fluctuated state.
The weakly correlated disorder and the glassy features of lattice heat capacity observed in the present system can be reasonably connected with the enhanced quantum fluctuation.
We believe that our discovery, as a macroscopic property of the quantum disordered state of phonons, has a significant impact on the rapidly growing research field of ``structural'' quantum criticality and permits us to go further into the unexplored fields.

\section{Conclusion}

We have discovered glass-like features at low temperature in terms of the lattice heat capacity of Ba$_{1-x}$Sr$_x$Al$_2$O$_4$, which is in a structurally disordered state outside the ferroelectric phase. 
In this state, the structural fluctuation associated with weakly correlated disorder is dominant as a consequence of the suppression of the M$_2$ and K$_2$ modes.
This weakly correlated disorder is responsible for the glass-like features observed in the low-temperature lattice heat capacity.
The structural disorder and the glassy heat capacity observed in the present system are considered one of the macroscopic properties exhibited by the quantum criticality of phonons.

\begin{acknowledgements}

We thank Prof. T. Doi and Prof. S. Horii (Kyoto University) for their support of compositional analysis using the inductively coupled plasma method.
This work was supported by a JSPS Grant-in-Aid for Scientific Research on Innovative Areas ``Mixed-anion'' (Grant Number 17H05487) and JSPS KAKENHI (Grant Number 17K14323).
The synchrotron radiation experiments were performed at BL02B1 of SPring-8 with the approval of the Japan Synchrotron Radiation Research Institute (JASRI) (Proposal No. 2017B1460).

\end{acknowledgements}
%\clearpage

\end{document}